\documentclass[12pt]{article}

\usepackage{geometry}                
\usepackage{graphicx}
\usepackage{amssymb}
\usepackage{epstopdf}
\usepackage[T1]{fontenc}
\usepackage{lmodern}
\usepackage{textcomp}
\usepackage{microtype}

\usepackage{amsmath}
\usepackage{amssymb}
\usepackage{subfigure}
\usepackage{eqlist} 
\usepackage[nosepfour,warning,np,debug,autolanguage]{numprint}
\usepackage{acro}
\usepackage{booktabs}

\DeclareGraphicsExtensions{.pdf, .jpg}
\usepackage{cite}
\usepackage{epsfig}
\usepackage{graphics}
\usepackage{lscape}
\usepackage{float}
\usepackage{multirow}
\usepackage[toc,page]{appendix}
\usepackage{setspace}
\usepackage{color}
\usepackage{url}
\usepackage{titlesec}

\newcommand{\beeq}{\begin{equation}}
\newcommand{\eeeq}{\end{equation}}
\newcommand{\be}{\begin{equation}}
\newcommand{\ee}{\end{equation}}

\newtheorem{theorem}{Lemma}

\newcommand{\RN}[1]{%
  \textup{\uppercase\expandafter{\romannumeral#1}}%
  }

\usepackage[linesnumbered,ruled,vlined]{algorithm2e}
\graphicspath{{Figures/}}

\usepackage{natbib}
 \bibpunct[, ]{(}{)}{,}{a}{}{,}%
\title{Registration-free localization of defects in 3-D parts from mesh metrology data using functional maps\vspace{-.5cm} }

\author{
{\small Xueqi Zhao}\\
{\small Enrique del Castillo\footnote{Corresponding author. Dr. Castillo is Distinguished Professor of Industrial \& Manufacturing Engineering and Professor of Statistics. e-mail: exd13@psu.edu}}\\  
{\small Engineering Statistics and Machine Learning Laboratory}\\
{\small Department of Industrial and Manufacturing Engineering and Dept. of Statistics}\\
{\small The Pennsylvania State University, University Park, PA 16802, USA}}\vspace{0.1cm}
\date{\small Version: December 2021}
\begin{document}
\maketitle\vspace{-1.5cm}
\begin{abstract}
Spectral Laplacian methods, widely used in computer graphics and manifold learning, have been recently proposed for the Statistical Process Control (SPC) of a sequence of manufactured parts, whose 3-dimensional metrology is acquired with non-contact sensors. These techniques  provide an {\em intrinsic} solution to the SPC problem, that is, a solution exclusively based on measurements on the scanned surfaces or 2-manifolds without making reference to their ambient space. These methods, therefore, avoid the computationally expensive, non-convex registration step needed to align the parts, as required by previous methods for SPC based on 3-dimensional measurements. Once a SPC mechanism triggers and out-of-control alarm, however, an additional problem remains: that of locating where on the surface of the part that triggered the SPC alarm there is a significant shape difference with respect to either an in-control part or its nominal (CAD) design. In the past, only registration-based solutions existed for this problem.  In this paper, we present a new registration-free solution to the part localization problem. Our approach uses a functional map between the manifolds to be compared, that is, a map between functions defined on each manifold based on intrinsic differential operators, in particular, the Laplace-Beltrami operator, in order to construct a point to point mapping between the two manifolds and be able to locate defects on the suspected part. A recursive partitioning algorithm is presented to define a region of interest on the surface of the part where defects are likely to occur, which results in considerable computational advantages. The functional map method involves a very large number of point-to-point comparisons based on noisy measurements, and a statistical thresholding method is presented to filter the false positives in the underlying massive multiple comparisons problem.
\end{abstract}



%




\section{Introduction}
Recent advances in  Statistical Process Control \citep{ZhaoEDC,ZhaoEDC-PE} permit the on-line monitoring of 3-dimensional (3D) parts scanned with non-contact sensors while avoiding registration of the parts, a non-convex, computationally expensive procedure that superimposes all parts such that they have the same location and orientation in the 3D space in which the parts are embedded. These advances are based instead on differential-geometric properties that are {\em intrinsic} to the surface (or 2-manifold) where metrology data occurs. Intrinsic differential properties of a manifold are those that can be totally computed from measurements on the manifold, and are independent of the ambient space. These properties can therefore be compared between manifolds {\em without} the need to register them first.

Once a part is detected to be the result of an out-of-control process condition, however, an additional problem remains: that of locating the {\em specific} defects on the surface of the part that has resulted in a significant shape difference with respect to either an in-control set of parts or the computer aided design (CAD) of the part. A part localization diagnostic that requires registration was suggested in \cite{ZhaoEDC}  based on the well-known Iterative Closest Point (ICP), which works for meshes with different number of points, an advantage over traditional statistical shape analysis methods based on Procrustes superposition which require an equal number of corresponding points in each object, an untenable requirement in non-contact sensed data.  ICP-based methods are computing intensive and based on a non-convex objective so they cannot guarantee a global optimum registration. Commercial inspection software use proprietary versions of the ICP algorithm or its many variants to map the parts and highlight deviations from nominal, in a process that requires considerable computing time that is not applicable for on-line diagnostics. The increasing availability of hand-held non-contact sensors in turn increases the desirability of a defect localization method that does not first need to place the parts on a common location and orientation.

Our goal, therefore, is to present an intrinsic solution to the defect localization problem which totally {\em avoids} the combinatorial registration problem while aligning the defective part and the CAD design (or an in-control part) and highlighting their statistically significant differences. The method, presented in section \ref{FM.FM}, uses the eigenvectors of the Laplace-Beltrami (LB)  operator, which captures local curvature properties of a surface. 

The following basic notions are used in the sequel of this paper.  The LB operator extends the notion of the Laplacian of  a real-valued function defined on flat Euclidean space to a function instead defined on a (possibly curved) manifold $\mathcal M \subset \mathbb{R}^n$, $f: \mathcal M \rightarrow \mathbb{R}$:
\be
\Delta_{\mathcal{M}} f = - div_{\mathcal{M}} \; \nabla_{\mathcal{M}} f
\ee
The LB operator is therefore the negative divergence of the gradient of the function, and it encodes not only the curvature of $f$, but also the curvature of the {\em manifold} $\mathcal M$ itself. For this reason, this operator is widely used in computer graphics and in manifold learning (e.g., see \cite{Belkin2008,reuter2006laplacebook,hamidian2019surface}). The LB operator appears in the heat diffusion partial differential equation \citep{Evans,ZhaoEDC}: 
\be \frac{\partial u(x,t)}{\partial t} = \Delta_{\mathcal{M}} u(x,t) \label{HeatEq} \ee
where $u(x,t)$ models the temperature at location $x \in \mathcal M$ at time $t$. By separation of variables, and considering only the spatial solution of the equation, one finds the so-called Helmholtz differential equation:
\[\Delta_{\mathcal{M}} \phi(x) = \lambda \phi(x)\]
The eigenvalues $0 \leq \lambda_1, \lambda_2,.... \uparrow + \infty$ define the spectrum of the LB operator, with corresponding 
eigenfunctions $\{\phi_k\}$. In practice, the manifold $\mathcal M$ is discretized and so is the LB operator, resulting in a matrix operator acting on vectors. The LB eigenvectors are sometimes used for segmentation purposes, since for a connected manifold the two {\em nodal domains} of the eigenfunction $\phi_2(x)$ associated with $\lambda_2$ divide $\mathcal M$  in geometrically meaningful ways, see \cite{ChavelBook}. For more on the LB operator and its spectrum see \cite{ZhaoEDC} and the many references therein. \cite{reuter2006laplacebook} shows how to more accurately estimate the LB operator using Finite Element Methods (FEM) based on a triangulation (or mesh) of a manifold, and \cite{ZhaoEDC-PE} utilize FEMs for implementing a non-parametric SPC control chart based on the spectrum of the LB operator for mesh and voxel data.

In this paper, a {\em functional map} is first constructed to establish the correspondence between two sets of LB eigenvectors, calculated from two parts, the potentially defective part and the nominal or acceptable part taken from an in-control operation period of the process. Then, for each point on the defective part, its best match on the CAD model or on the in-control part can be found based on the functional map. Finally, the shape dissimilarity between each point and its best match is calculated using intrinsic geometrical properties, and regions with high dissimilarities are highlighted as local defects. The  computational complexity of the new method is studied in section 4. Since the functional mapping method involves therefore a very large number of comparisons, a thresholding method is presented in section 5 to consider not only large but statistically significantly large deviations from nominal, and to filter the false positives in what constitutes an underlying classical multiple comparisons problem. We finally show in section 6 how to adapt the method to consider a user-defined region of interest (ROI) on the part, resulting in considerable computational savings, and present an iterative intrinsic segmentation algorithm that identifies a ROI in case it can not be defined a priori by the user. The paper concludes with a summary of findings and directions for further research, and the Appendix contains further technical discussion.

\section{Defining a functional map between two manifolds} \label{FM.FM}
Given two manifolds $\mathcal{A}$ and $\mathcal{B}$ of similar but not necessarily equal geometrical shape, we wish to find points or regions on $\mathcal{B}$ corresponding to given points or regions of interest on $\mathcal{A}$ while avoiding the combinatorial complexity of registration methods. A likely scenario of this problem occurs in SPC where $\mathcal A$ represents a noisy scanned part and $\mathcal B$ represents a noise-free mesh giving the nominal CAD design of the part. We want to map any differences between $\mathcal A$ and $\mathcal B$ and display the differences in order to better diagnose the manufacturing process. In industrial practice, and contrary to computer graphics where similar problems exist in shape classification, the differences we wish to detect are small, not always perceivable to the human eye, and buried in measurement and manufacturing noise. 

\begin{figure}[h]
	\begin{center}
		\includegraphics[scale=0.5]{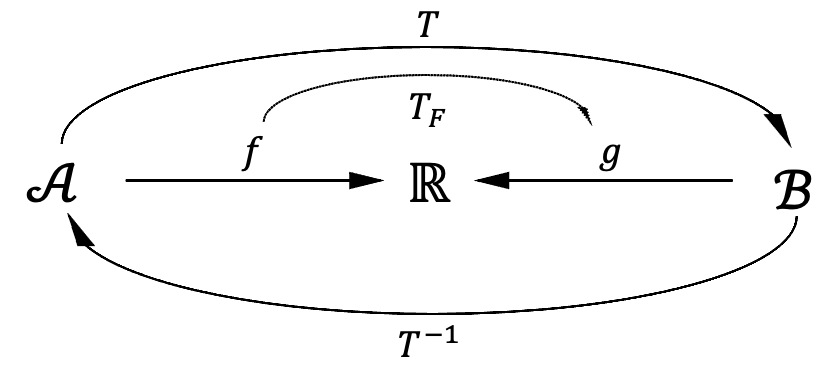}
		\caption{The relation between $\mathcal A$, $\mathcal B$, $\mathbb{R}$ , $T$, $T^{-1}$, and $T_F$.}
	\label{FuncMapDiagram}
	\end{center}
\end{figure}

Given a point-to-point mapping $T: \mathcal{A}\to\mathcal{B}$ one could highlight $T(P)\subseteq\mathcal{B}$ whenever a set of points $P\subseteq\mathcal{A}$ are selected. We do not attempt finding $T$ directly (a combinatorial problem), and we instead construct a {\em functional map}  \citep{ovsjanikov2012functional} and use it to find the mapping $T$. Assume for a moment $T$ is known. Then, for any scalar function on $\mathcal{A}$, $f: \mathcal{A}\to\mathbb{R}$, we could obtain a scalar function on $\mathcal{B}$, $g: \mathcal{B}\to\mathbb{R}$ via $g=f\circ T^{-1}$, assuming $T$ is bijective. The correspondence between all such pairs of functions can be represented by the functional map (or map between functions): 
\[ T_F: \mathcal{F}(\mathcal{A},\mathbb{R})\to\mathcal{F}(\mathcal{B},\mathbb{R}),\] 
where $\mathcal{F}(\mathcal{W},\mathbb{R})$ is the space of real-valued functions on manifold $\mathcal{W}$. Since $T_F(f)=f\circ T^{-1}$, this functional map depends only on the original point mapping $T$, and we say $T_F$ is induced by $T$ (see Figure \ref{FuncMapDiagram}). Importantly, $T$ can be easily recovered point by point once $T_F$ is known, by defining
\beeq
\label{findT}
T(x)=\operatorname*{argmax}_y ~T_F(\delta_x)(y)
\eeeq
where $\delta_x$ is the indicator function at point $x\in\mathcal{A}$. This equation holds because $T_F(\delta_x)(f)=\delta_x\circ T^{-1}(y)$ is nonzero only when $T^{-1}(y)=x$, in other words, when $y=T(x)$. Therefore, we can determine the point mapping $T$ by finding the functional mapping $T_F$ first.

Suppose now $\mathcal{F}(\mathcal{A},\mathbb{R})$ and $\mathcal{F}(\mathcal{B},\mathbb{R})$ are each equipped with orthonormal basis functions $\{\phi_i^A\}_i$ and $\{\phi_j^B\}_j$, respectively. Then for any $i$, $T_F(\phi_i^A)\in \mathcal{F}(\mathcal{B},\mathbb{R})$ can be expressed as a unique linear combination of $\{\phi_j^B\}_j$:
\be
T_F(\phi_i^A) = \sum_j c_{ij}\phi_j^B
\ee
where as Lemma \ref{LemmaC} in the Appendix shows, all we need to find is therefore the coefficients $\{c_{ij}\}$.

Now we discuss how to determine the $C$ matrix containing the $\{c_{ij}\}$ coefficients. If we available have $K$ pairs of functions $(f_k, g_k)$, $k=1, 2, ..., K$, such that $f_k\in\mathcal{F}(\mathcal{A},\mathbb{R}), \forall k=1, 2, ..., K$, $g_k\in\mathcal{F}(\mathcal{B},\mathbb{R}), \forall k=1, 2, ..., K$, and $T_F(f_k)=g_k, \forall k=1, 2, ..., K$, then both $f_k$ and $g_k$ can be represented by their corresponding basis, namely
$$f_k=\sum_i \alpha_{ik} \phi_i^A \quad \mbox{and} \quad g_k=\sum_j \beta_{jk} \phi_j^B.$$
From expression (\ref{Tf}) in the Appendix (Lemma \ref{LemmaC}) we have that
$$g_k=T_F(f_k)=\sum_j\left(\sum_i \alpha_{ik}c_{ij}\right)\phi_j^B.$$
Since the basis functions $\{\phi_j^B\}_j$ are linearly independent, we obtain $\beta_{jk}=\sum_i \alpha_{ik}c_{ij}$, $\forall j$, which can be organized in matrix form as
\beeq
{\boldsymbol \beta}_k\triangleq\begin{pmatrix} \beta_{1k} \\ \beta_{2k} \\ \vdots \\ \beta_{jk} \\ \vdots \end{pmatrix}=\begin{pmatrix}
c_{11} & c_{21} & \cdots & c_{i1} & \cdots \\
c_{12} & c_{22} & \cdots & c_{i2} & \cdots \\
\vdots & \vdots & \ddots & \vdots & \cdots \\
c_{1j} & c_{2j} & \cdots & c_{ij} & \cdots \\
\vdots & \vdots & \vdots & \vdots & \ddots \end{pmatrix}
\begin{pmatrix} \alpha_{1k} \\ \alpha_{2k} \\ \vdots \\ \alpha_{ik} \\ \vdots \end{pmatrix}\triangleq C'{\boldsymbol \alpha}_k
\eeeq
Given the $K$ pairs of function values we can define matrices $B=\begin{pmatrix} {\boldsymbol \beta}_1 & {\boldsymbol \beta}_2 & \cdots & {\boldsymbol \beta}_K\end{pmatrix}$ and  $A=\begin{pmatrix} {\boldsymbol \alpha}_1 & {\boldsymbol \alpha}_2 & \cdots & {\boldsymbol \alpha}_K\end{pmatrix}$. Then we have
\beeq
\label{findC}
B=C'A.
\eeeq
Therefore, given a basis in $\mathcal{F}(\mathcal{A},\mathbb{R})$ and in $\mathcal{F}(\mathcal{B},\mathbb{R})$, and $K$ pairs of corresponding functions $(f_k, g_k)$, matrix $C$ can be calculated by solving eq (\ref{findC}), which completely determines $T_F$ and can be used to recover the point matching $T$ as desired. The whole procedure is summarized in the following algorithm:\\
\begin{algorithm}[H]
\SetAlgoLined
\KwIn{Manifolds (surface meshes) $\mathcal{A}$ and $\mathcal{B}$}
\KwOut{Point-to-point mapping $T: \mathcal{A}\to\mathcal{B}$}
Find a set of orthonormal basis for $\mathcal{F}(\mathcal{A},\mathbb{R})$, $\{\phi_i^A\}_i$\\
Find a set of orthonormal basis for $\mathcal{F}(\mathcal{B},\mathbb{R})$, $\{\phi_j^B\}_j$\\
Find $K$ pairs of known correspondences $(f_k, g_k)$, $k=1, 2, ..., K$\\
Calculate the entries in the $A$ matrix by $\alpha_{ik}=\langle f_k, \phi_i^A\rangle$\\
Calculate the entries in the $B$ matrix by $\beta_{jk}=\langle g_k, \phi_j^B\rangle$\\
Solve for matrix $C$ in equation $B=C'A$\\
\For{each point $x\in\mathcal{A}$}{
	Define $\delta_x\in\mathcal{F}(\mathcal{A},\mathbb{R})$ to be the indicator function of $x$\\
	Find the image of $\delta_x$ by $T_F(\delta_x)=\sum_j\left(\sum_i \alpha_{i}^xc_{ij}\right)\phi_j^B$, where $\alpha_{i}^x=\langle \delta_x, \phi_i^A\rangle$\\
	Find the image of $x$ by eq (\ref{findT}): $T(x)=\operatorname*{argmax}_y T_F(\delta_x)(y)$
}
\caption{The functional map framework}
\label{algo1}
\end{algorithm}

\section{Implementation of the functional map algorithm} \label{FM.Implement}

Specific choices are needed to implement Algorithm \ref{algo1}. In particular, choices are needed for the orthonormal basis and for the functions used to find  correspondences. We discuss next these choices, as well as a modified method to recover $T$ in the presence of noise. 

\subsection{Selection of orthonormal basis functions} \label{FM.basis}
The eigenfunctions of the LB operator provide a natural orthonormal basis of each manifold for steps 1 and 2 in Algorithm \ref{algo1}. 
Numerically, we work with an estimated LB operator obtained from a mesh of connected measurements, and the resulting eigenvectors are the discretized version of the LB eigenfunctions, both of which are intrinsic and therefore independent of rigid transformations. 

\begin{figure}[h]
	\centering
		\includegraphics[width=\textwidth]{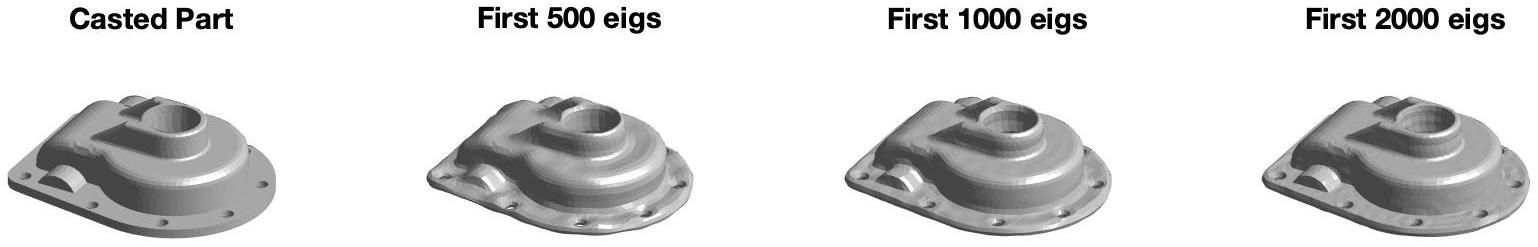}
	\caption{Approximated meshes of a casted part using the first 500, 1000, and 2000 eigenvectors, respectively.}
	\label{recover}
\end{figure}

Recall that by convention, the eigenvalue-eigenfunction pairs are ordered such that the eigenvalues are non-decreasing. The eigenfunctions associated with the leading eigenvalues describe the general shape of the manifold, while eigenfunctions associated with eigenvalues that appear later include smaller shape features. Since the eigenfunctions form the basis of the eigenspace corresponding to their eigenvalues, the eigenfunctions contain multi-scale shape information \citep{SunHKS2009}. Such hierarchical relationship means that a large portion of the function space can be represented by the first several eigenfunctions. Figure \ref{recover} shows a casted part with around 5000 points (leftmost display) and its approximations using increasing numbers of eigenvectors. Although the maximum number of eigenvectors one can use is the same as the mesh size (about 5000 in this example), we are able to capture the general shape of the part with only the first 500 eigenvectors, with smaller details appearing as more eigenvectors are considered. This indicates that in the first two steps of Algorithm \ref{algo1}, we can use the ordered eigenvectors up to a certain maximum index $p$, instead of the complete set of basis, which reduces both the computational and the storage costs.

Another advantage of using the LB eigenfunctions as the basis comes from their correspondence across objects with similar shapes. For two manifolds with similar or almost identical shapes, their eigenfunctions under the same index are likely to contain the same shape information and therefore correspond to each other under the functional mapping $T_F$. When the parts have no symmetries in their shape, no eigenvalue is repeated, and each eigenfunction represents an eigenspace of dimension 1. In this case, the eigenfunctions correspond exactly with each other under the same index, so matrix $C$ is diagonal. This greatly reduces the number of unknowns in matrix $C$ from $p\times p$ to $p$, and equation (\ref{findC}) reduces to
\beeq
\label{findC_simple}
\beta_{jk}=c_{jj}\alpha_{jk}, ~\forall j, k
\eeeq
When working with metrology data from actual manufactured parts, both $\beta_{jk}$ and $\alpha_{jk}$ will be calculated with errors, so for a given $j$, equation (\ref{findC_simple}) should be seen as a simple linear regression model where $c_{jj}$ is to be estimated based on $K$ observations rather than accurately calculated through a system of equations. Least squares and ridge regression can be used for estimation as discussed in more detail in the Appendix.

In practice, the manifold is approximated by a triangulation mesh. Consequently, a discrete LB operator is calculated in the form of a matrix and the eigenvectors are used to approximate the eigenfunctions. Based on results by \cite{ZhaoEDC-PE}, we use a cubic FEM approach to numerically obtain the LB operator in this paper. We point out that the LB eigenvectors of the discrete LB operator approximations do not form an orthogonal basis with the Euclidean inner product \citep{rustamov2007}, so the Gram-Schmidt process should first be applied to orthonormalize the eigenvectors in Algorithm \ref{algo1}. 

\subsection{Defining known correspondences between manifolds}
\label{FM.HKS}
For the functions needed in order to find the correspondences in step 3 of Algorithm \ref{algo1} it is convenient to use functions intrinsically defined on each manifold. For this reason we propose to use the so-called {\em heat kernel signature} (HKS) function \citep{SunHKS2009} on each manifold. 
The {\em heat kernel} $k_t(x,y)$ appears in the fundamental space-time solution of the heat difussion partial differential equation and is defined for $x,y \in \mathcal M$ as $k_t(x,y) = \sum_{i=0}^{\infty}  e^{\lambda_i t} \phi_i(x) \phi_i(y)$ \citep{Evans,ZhaoEDC}. It represents the amount of heat transmitted from point $x$ to point $y$ on the mesh by time $t$ if at $t=0$ there is a unit heat source at $x$.  As such, it is invariant with respect to rigid transformations. The HKS is then defined as $k_t(x,x)$ and describes the shape information around point $x$ within time $t$. Although $t$ stands for time, it can also represent distance here, since a larger amount of time allows heat to travel further. 

Since $\{\lambda_i\}$ is non-decreasing, as $t$ increases, $k_t(x,x)$ is dominated by the $\lambda_0$ term and approaches the limit $\phi_0(x)^2=1/n$ for all $x$ on a mesh, with $n$ being the mesh size. \cite{SunHKS2009}  suggest scaling the HKS by its integral over the whole manifold $\mathcal A$ to make differences at different times contribute approximately equally. They call this the {\em scaled HKS}, defined as:
\beeq
\text{scaled } k_t(x,x) = \frac{k_t(x,x)}{\int k_t(x,x)dx}
=\frac{\sum_{i=0}^{\infty} e^{-\lambda_i t} \phi_i(x)^2}{\sum_{i=1}^{\infty} e^{-\lambda_i t}}
\label{Scaled HKS}
\eeeq
The limit of the scaled HKS as $t\to\infty$ is different from that of the original HKS, yet still proportional to $1/n$. To make the scaled HKS comparable across different meshes, we further normalize the scaled HKS by $n$, the mesh size. In the Appendix we illustrate how this normalization is not only intrinsic but also independent of the mesh size, an important matter when making comparisons between triangulations of manifolds which will most likely not be of the same size.
Therefore, given $t$, we treat the normalized and scaled HKS as a function defined on manifold $\mathcal A$, that is
\beeq
\label{scaledHKS}
f_t(x)=\text{scaled } k_t(x,x) \cdot n_A
\eeeq
thus $f_t\in\mathcal{F}(\mathcal{A},\mathbb{R})$. Similarly, we define $g_t\in\mathcal{F}(\mathcal{B},\mathbb{R})$ to be the normalized and scaled HKS on manifold $\mathcal B$. Since time $t$ is a physical parameter, $f_t$ naturally corresponds to $g_t$ through $g_t=T_F(f_t)$ for any given $t$, providing a pair of observations for the regression model (\ref{findC_simple}). 

To better estimate matrix $C$ in step 6, we want to have a wide range of $t$ values such that the variation in the normalized and scaled HKS is captured as completely possible, especially for small $t$. Therefore, for all results presented in this paper, we use 100 values of $t$ logarithmically sampled from $t_{\min}=4\log10/\lambda_p$ to $t_{\max}=4\log10/\lambda_1$. The justification for the selection of this range is explained in the Appendix.

\subsection{Recovering the point-to-point transformation $T$}
\begin{figure}
	\centering
		\includegraphics[width=\textwidth]{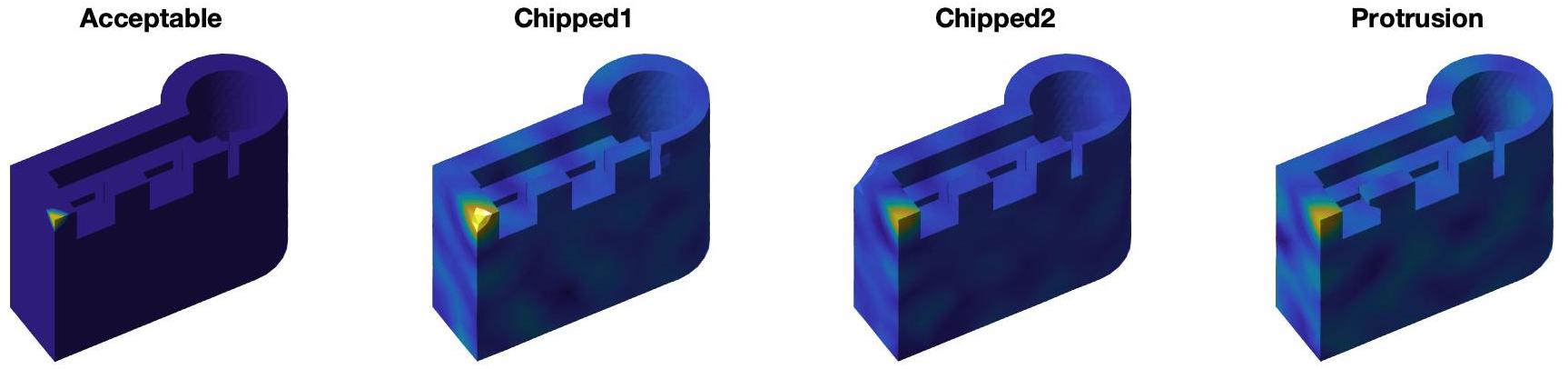}
	\caption{Defective parts (right most three parts) color-coded by $T_F(\delta_x)$ when $x$ is fixed at the yellow corner on the acceptable part (first part on the left).}
	\label{plotTf}
\end{figure}

In step 10 of Algorithm \ref{algo1},  $T_F(\delta_x)(\cdot)$ is calculated, in practice, in the presence of noise, which prevents it from being a perfect indicator function taking value 1 at point $T(x)$ and 0 everywhere else, as expected in the noise-free case. Instead, it is simply a real-valued function taking different values at different points, as shown in Figure \ref{plotTf}, where the three defective parts in Figure \ref{plotTf} are color-coded according to $T_F(\delta_x)(\cdot)$ with $\delta_x$ being the indicator function of the highlighted point on the acceptable part on the left. Note how all three defective parts have varying colors, indicating fluctuating function values due to noise. This is why we set $T(x)$ equal to the maximum of $T_F(\delta_x)(\cdot)$, rather than $T_F(\delta_x)(\cdot)=1$.

Given that noise in $T_F(\delta_x)(\cdot)$ may slightly shift its maximizer around the true $T(x)$, in step 10 in our Algorithm we consider the top $m$ points with the highest function values of $T_F(\delta_x)(\cdot)$ and set the point that is the ``closest'' to $x$ to be $T(x)$. From our experiments with meshes of around 2000 points, $m=5$ to $10$ was enough to ensure a high quality point matching. In general, $m$ needs to increase accordingly when manifold $\mathcal B$ becomes denser. 

Next, we need to measure the closeness between two points from different manifolds, in order to determine if the two points correspond to each other. In registration methods, this is usually measured by the Euclidean distance between these two points since the two objects have been brought to the same orientation and position in the ambient space. Instead, we propose to use the difference between the normalized and scaled HKS, treated as a function of $t$ given the point, to evaluate the similarity between two points $x\in\mathcal A$ and $y_i\in\mathcal B$, where $T_F(\delta_x)(y_i)$ gives the $i$th highest value for function $T_F(\delta_x)(\cdot)$, $i=1, ..., m$. 

As mentioned above, the normalized and scaled HKS $f_t(x)$ codifies the local shape information around $x$ and is independent of mesh qualities and rigid transformations. Therefore, for any $t$, $g_t(y)$ should become closer to $f_t(x)$ if and only if $y\in\mathcal B$ is closer to the true $T(x)$ for $x\in\mathcal A$. We take advantage of this property and propose to use the following instead of equation (\ref{findT}):
\beeq
\label{findTnew}
T(x)=\operatorname*{argmin}_{y_i, i=1, ..., m} \sqrt{\sum_t(f_t(x)-g_t(y_i))^2}
\eeeq
with $y_i$ is defined above. Defining the vectors of normalized and scaled HKS at different $t$: 
$
\bf f_x = \begin{pmatrix} f_{t_1}(x) & f_{t_2}(x) & \cdots & f_{t_{100}}(x) \end{pmatrix},$ and $
\bf g_{y_i} = \begin{pmatrix} g_{t_1}(y_i) & g_{t_2}(y_i) & \cdots & g_{t_{100}}(y_i) \end{pmatrix},
$
then the new criterion simply minimizes the norm of the difference between the HKS vectors:
\beeq
\label{findTnew2}
T(x)=\operatorname*{argmin}_{y_i, i=1, ..., m}
\|\bf f_x-\bf g_{y_i} \|.
\eeeq

\subsection{Implementation summary}

The complete implementation of Algorithm \ref{algo1} is listed in Algorithm \ref{algo2}. Here $p$ and $K$ are the hyper-parameters that should be set by the user based on the mesh qualities. Intuitively, these parameters should increase for larger meshes to provide the algorithm with more details about the two shapes. For all examples in this paper, $p=200$ and $K=100$ were used. All quantities used are related to the heat diffusion process and are therefore intrinsic, making the registration step unnecessary. We apply the Gram-Schmidt process on the LB eigenvectors to obtain the orthonormal basis of the space of real-valued functions, use the normalized and scaled HKS at a sequence of $t$ values as known corresponding functions, and recover the point matching $T$ by minimizing the difference in this HKS.  
\begin{algorithm}[h]
\SetAlgoLined
\KwIn{Defective part as mesh $\mathcal{A}$ and CAD design as mesh $\mathcal{B}$}
\KwOut{Point-to-point mapping $T: \mathcal{A}\to\mathcal{B}$}
Estimate the first $p$ LB eigenvalues $\{\lambda_i^A\}$ and eigenvectors $\{\phi_i^A\}$ of $\mathcal{A}$ using the cubic FEM method\\
Estimate the first $p$ LB eigenvalues $\{\lambda_i^B\}$ and eigenvectors $\{\phi_i^B\}$ of $\mathcal{B}$ using the cubic FEM method\\
Apply the Gram-Schmidt process to obtain the orthonormalized eigenvectors $\{\tilde{\phi}_i^A\}$ and $\{\tilde{\phi}_i^B\}$\\
Calculate the normalized and scaled HKS as defined in equation (\ref{scaledHKS}), $f_t$ and $g_t$, for $\mathcal{A}$ and $\mathcal{B}$, respectively, with $K$ values of $t$ logrithmically sampled from $4\log10/\lambda_p$ to $4\log10/\lambda_1$ \\
Calculate the entries in the $A$ matrix by $\alpha_{ik}=\langle f_k, \tilde{\phi}_i^A\rangle$, $i=1, ..., p$, $k=1, ..., K$\\
Calculate the entries in the $B$ matrix by $\beta_{jk}=\langle g_k, \tilde{\phi}_j^B\rangle$, $i=1, ..., p$, $k=1, ..., K$\\
Calculate matrix $C$ using the least squares method (details discussed in the Appendix)\\
\For{each point $x\in\mathcal{A}$}{
	Define $\delta_x\in\mathcal{F}(\mathcal{A},\mathbb{R})$ to be the indicator function of $x$\\
	Find the image of $\delta_x$ by $T_F(\delta_x)=\sum_j\left(\sum_i \alpha_{i}^xc_{ij}\right)\tilde{\phi}_j^B$, where $\alpha_{i}^x=\langle \delta_x, \tilde{\phi}_i^A\rangle=\tilde{\phi}_i^A(x)$\\
	Find the image of $x$ by equation (\ref{findTnew2})
}
\caption{The functional map implementation}
\label{algo2}
\end{algorithm}

\begin{figure}
	\centering
		\includegraphics[scale=0.15]{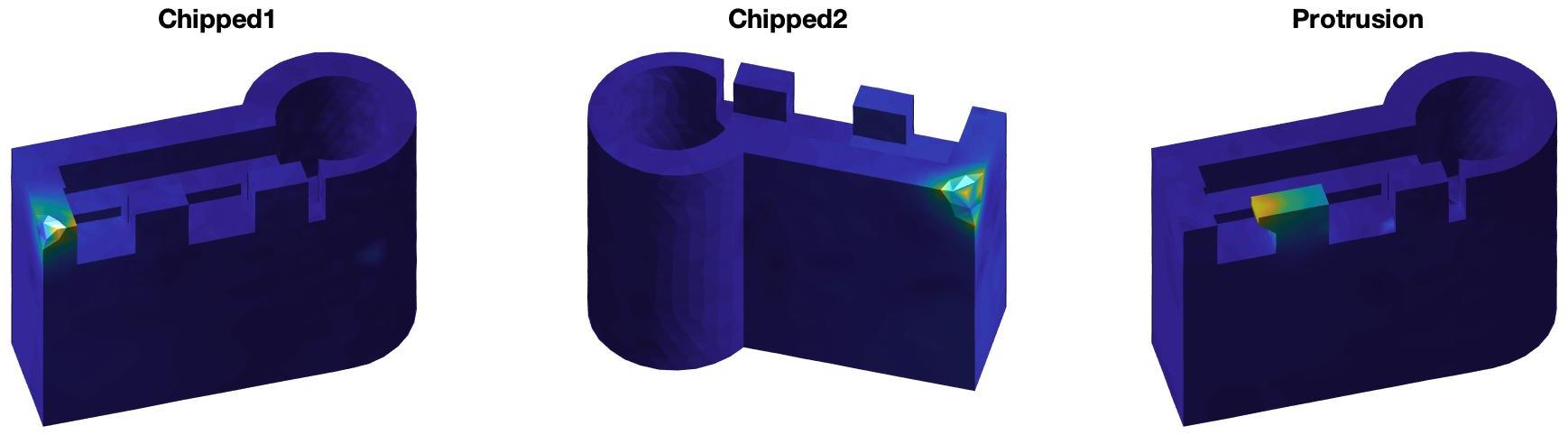}
	\caption{Defective parts (as manifold $\mathcal A$) color-coded by $\|\bf f_x-\bf g_{T(x)} \|$ for all $x\in \mathcal A$. $\mathcal B$ is the CAD model (left graph in Figure \ref{recover}). All local defects are correctly highlighted with our functional mapping method. $m=5$.}
	\label{highlight}
\end{figure}

When performing on-line SPC, the part that triggers an out-of-control alarm and needs to be examined is set as manifold $\mathcal A$ and the CAD model or an acceptable part (from a reference set of in-control parts) acts as manifold $\mathcal B$. Once the point-to-point mapping $T$ is recovered, each point $x\in\mathcal A$ is associated with a value that measures how this location in $\mathcal A$ deviates from manifold $\mathcal B$, a deviation calculated by $\|\bf f_x-\bf g_{T(x)} \|$. Points on the mesh with high deviations are labeled as belonging to a defective region. For example, Figure \ref{highlight} displays the three defective parts in Figure \ref{plotTf} color-coded by $\|\bf f_x-\bf g_{T(x)} \|$ with lighter colors corresponding to higher values. The local defect area in all three parts is correctly highlighted, indicating the remarkable performance of the functional map method.

\section{Properties of the functional map method: computational complexity} \label{FM.Properties}

The functional map method has several desirable characteristics. In the Appendix we discuss how, compared to ICP-methods, it assures a global optimal point to point matching. We also discuss how the HKS we use is not only an intrinsic function but also is independent of the mesh sizes used. Here we discuss the computational complexity of the method.

\subsection{Computational Complexity}\label{FM.Complexity}
The computational complexity of the main steps in our functional mapping method is summarized in Table \ref{compute}. Since we choose $p=200$, $K=100$, $m=5$ for meshes of size around $n=1700$, the dominant term is $\mathcal{O}(n_An_Bp)\approx \mathcal{O}(n_An_B)$ as in the last row, resulting from step 10 in Algorithm \ref{algo2} to recover the point-to-point mapping $T$ given matrix $C$ (this can be further reduced when a particular region of interest is available, discussed later in Section \ref{FM.ROI}). As mentioned in  \cite{ovsjanikov2012functional}, this step can be more efficiently completed by finding the nearest neighbors in $\tilde{\Phi}^B$ for every point or row in $\tilde{\Phi}^AC$, where $\tilde{\Phi}^A$ and $\tilde{\Phi}^B$ are matrices consisting of the orthornormalized eigenvectors $\{\tilde{\phi}_i^A\}$ and $\{\tilde{\phi}_i^B\}$ as columns, respectively.  \cite{ovsjanikov2012functional} indicate that an efficient search algorithm can reduce the computational complexity from $\mathcal{O}(n_An_B)$ to $\mathcal{O}(n_A\log n_A+n_B\log n_B)$. The ICP algorithm, on the other hand, has a typical computational complexity of $\mathcal{O}(n_rn_An_B)$ for global matching, where $n_A$ and $n_B$ are the mesh sizes for $\mathcal A$ and $\mathcal B$, respectively, and $n_r$ is the number of initial rotation states to avoid local optima \citep{Besl}. In practice, the mesh sizes $n_A$ and $n_B$ can be in the many tens of thousands, compared to which both $p$ and $n_r$ are negligible, so the computational complexity of both methods can be simplified to $\mathcal{O}(n_An_B)$.

\begin{table}[h]
\begin{center}
\begin{tabular}{c|c}
\hline
Main Steps & Computational Complexity \\
\hline	
Constructing the FEM LB matrices & $\mathcal{O}(n_A)+\mathcal{O}(n_B)$ \\
Solving for the first $p$ eigenvalues and eigenvectors & $\mathcal{O}(n_Ap)+\mathcal{O}(n_Bp)$ \\
Applying the Gram-Schmidt process & $\mathcal{O}(n_Ap^2)+\mathcal{O}(n_Bp^2)$ \\
Calculating the normalized and scaled HKS & $\mathcal{O}(n_ApK)+\mathcal{O}(n_BpK)$\\
Constructing the $A$ and $B$ matrices & $\mathcal{O}(n_ApK)+\mathcal{O}(n_BpK)$ \\
Calculating matrix $C$ & $\mathcal{O}(pK)$\\
Recovering the point-to-point matching $T$ & $\mathcal{O}(n_Bp+n_An_Bp+n_AmK)$\\
\hline
Total & $\mathcal{O}(n_An_Bp)$\\
\hline
\end{tabular}
\caption{Computational complexity of our functional mapping method. $n_A$ and $n_B$ are the mesh sizes for $\mathcal A$ and $\mathcal B$, respectively, $p$ is the number of eigenvalue-eigenvector pairs, $K$ is the number of $t$ values (known correspondences), and $m$ is the number of $y$'s considered to find $T(x)$.}
\label{compute}
\end{center}
\end{table}

\begin{figure}[h]
	\centering
		\includegraphics[width=\textwidth]{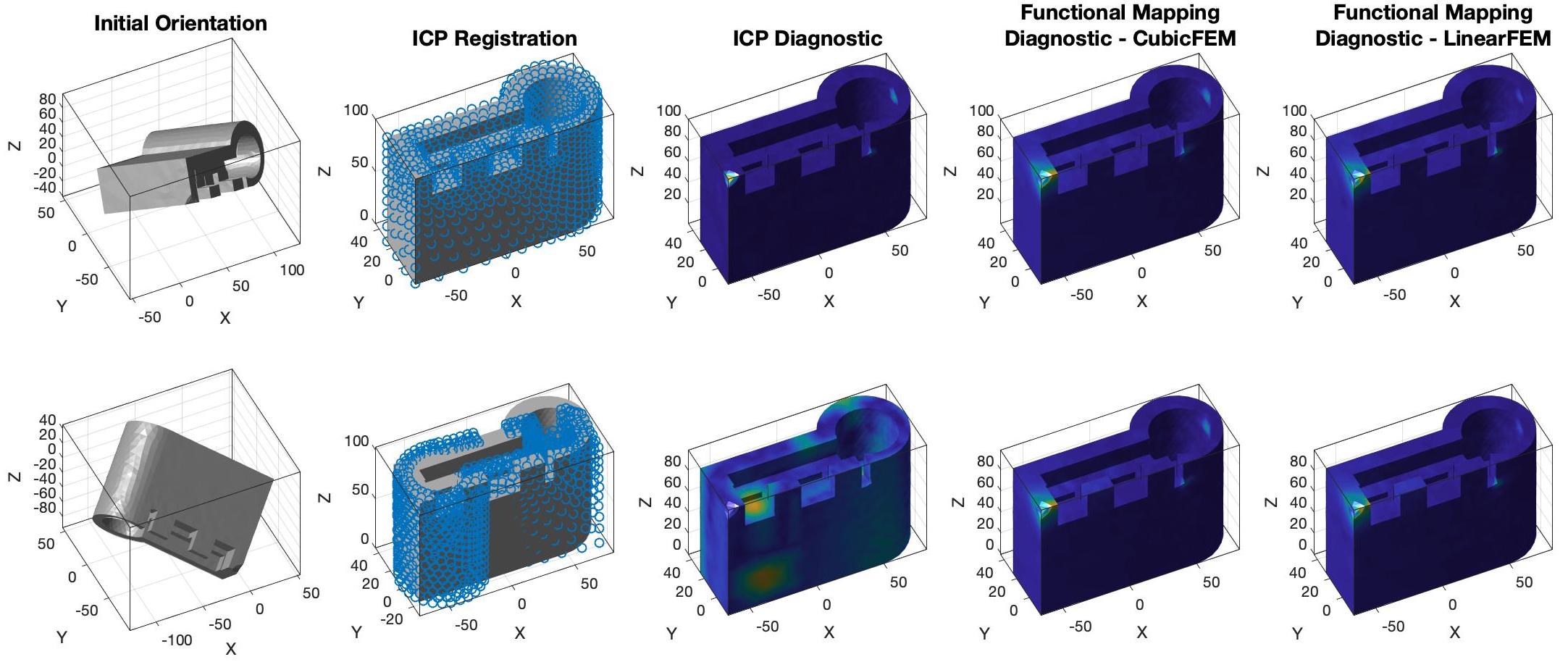}
	\caption{Examples of two different initial orientations of the ``chipped1'' part to be matched with an acceptable part. The ICP registration succeeds in the example on the first row but fails in the example  in the second row, while the functional map method correctly highlights the local defect area for both scenarios with both the cubic and the linear FEM. With increasing use of hand-held scanners in industry, widely different orientations are possible, and hence this is a desirable advantage of the functional map method.}
	\label{vsICP}
\end{figure}

\begin{figure}[h]
	\centering
		\includegraphics[width=\textwidth]{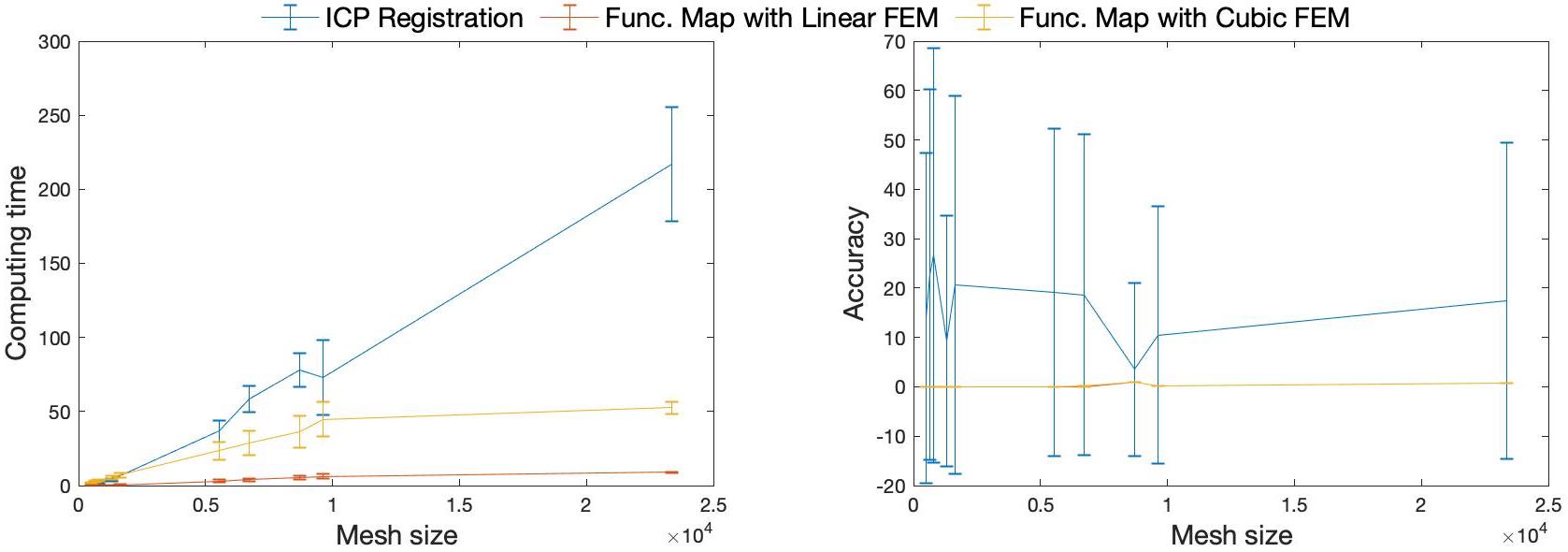}
	\caption{On the left, computing time (in seconds) comparisons and on the right, accuracy comparisons between the ICP registration method and the functional map method, as the mesh size increases (both linear and cubic FEM were evaluated for the functional map). Each scenario was simulated 100 times. The line plots represent the average and the length of the error bars are twice the standard deviation. The error bars for the two functional map methods on the right figure are too small to see compared to those for the ICP registration method. The plots representing the two functional map methods overlap due to their similar performance and the zoom out effect to include the ICP algorithm.} 
	\label{Time}
\end{figure}

Figure \ref{Time} compares the performance of ICP registration with that of our functional map method under both the linear and cubic FEM LB estimates, respectively. We select ten different mesh sizes ranging from 500 to 25,000 points, and run 100 replications for each combination of the algorithm and mesh size. All experiments were tested on the same computer with 4 GHz Quad-Core Intel Core i7 and 16G RAM. Both the average and standard deviation are used to summarize the results. The left figure plots the computing time of the three methods. As can be seen, regardless of the degree of the FEM method used, the functional map approach scales much better with the mesh size than the ICP method. 
The right display on Figure \ref{Time} compares the accuracy of the three methods, where we define
\beeq
\label{AccuracyMeasure}
\text{Accuracy} = \sum_{i=1, ..., n_A} \| T(x_i)-G(x_i) \|/n_A
\eeeq
where the $x_i$'s are the points on manifold $\mathcal{A}$ of size $n_A$, $T(x_i)$ is the best matching point for $x_i$ found on manifold $\mathcal{B}$ by the method being evaluated, and $G(\cdot)$ is the known ground truth mapping. Note the exceptionally large variability of the ICP algorithm, caused by bad initial orientations and local optima, as illustrated in Figure \ref{vsICP}. This makes the application of statistical methods to identify significant defects in ICP difficult. On the other hand, thanks to its stable behavior, the functional map scheme can be further equipped with multiple comparison hypothesis testing methods to account for manufacturing and measuring noise and to control the false positives, as discussed next.

\section{A thresholding method for the underlying multiple comparisons problem}  \label{FM.MultiComp}

In this section we propose an additional algorithm to better deal with measurement and manufacturing noise, in order to locate the region on the defective part whose deviation is not only the largest but also the most statistically significant. This implies statistical tests for all individual points, which leads to a multiple comparison problem. We introduce a single threshold method adapted from \cite{holmes1996nonparametric} to control the family-wise error rate. 

\begin{figure}[h]
	\centering
		\includegraphics[width=\textwidth]{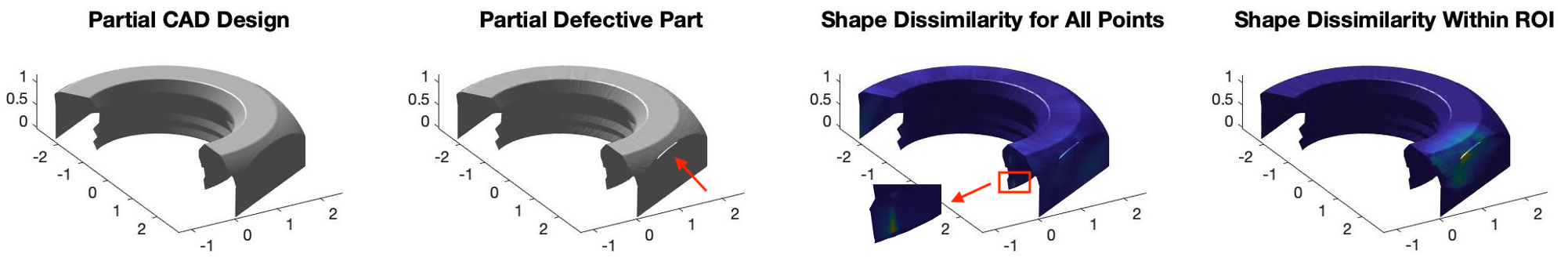}
	\caption{A partial defective hex nut with an indentured crack (marked by the red arrow in the second part from the left) is matched with its CAD design cut in the same way (first part on the left). A small region near the boundary (marked by the red box and enlarged to allow visualization) is falsely identified as having the highest dissimilarity against the CAD when all points on the partial mesh are considered (third part). The crack is correctly highlighted when we focus on a specific region of interest (fourth part), see Section \ref{FM.ROI}.}
	\label{highlightROI_zoom}
\end{figure}

In a manufacturing environment, metrology usually includes noise which is the sum of  manufacturing and measurement errors, which, if large, may cause difficulties to the functional map as presented thus far. For example, in Figure \ref{vsICP}, there is a small light blue area in the cylindrical region of the part in all five graphs where the local defects are correctly identified (the last three parts displayed on the first row and the last two parts on the second row). This area with relatively large deviations, indicated by its lighter colors, is caused by noise but detected by all three diagnostic algorithms. These false positives should be avoided by a robust diagnostic algorithm. To account for such noise as well as for the variability resulted from the unequal mesh sizes with non-corresponding points, we consider conducting hypothesis testing to differentiate the true local defect area from deviations caused by noise. 

In the general routine, a test statistic is calculated for each point and compared against a certain threshold, where only points with statistics exceeding the threshold are classified as significant. Usually the threshold is chosen to be the $100(1-\alpha)$th percentile of the null distribution to obtain a significance level of $\alpha$ for individual tests, so that the probability of falsely detecting a non-significant point is $\alpha$. However, with such point-wise hypothesis tests applied to all points on the mesh simultaneously, we expect $100\alpha\%$ of the points to be falsely detected as having significant deviations when none of them is actually significant. This is known as the multiple comparison problem, which leads to more points declared as significant than those that are truly significant, making the testing procedure less efficient. The most intuitive adjustment we can make to correct such problem and control the overall type I error, the probability of false positives, is by choosing a more strict threshold. 

There are various well-known methods in the literature that handle the multiple comparison problem (see, e.g., \cite{holm1979simple,nichols2012multiple, nichols2003controlling, genovese2002thresholding}). 
%
Due to its simplicity and performance, we adapted the single threshold test proposed by \cite{holmes1996nonparametric}, which was originally developed in neuroimaging.
If $D(x)\triangleq\|\bf f_x-\bf g_{T(x)} \|$ is the point-wise deviation from the CAD model for each point $x$ on the defective part, now we wish to evaluate the {\em statistical significance} of a test for $H_0: D(x)=0$ instead of arbitrarily highlighting differences of certain magnitude of $D(x)$ disregarding noise. Here a measure of the natural variability of $D(x)$ can be obtained by applying the functional map method (Algorithm \ref{algo2}) to parts produced while the process was in a state of statistical control (sometimes referred to as ``Phase I" operation of a manufacturing process). 

Suppose we have available $m_0$ Phase I parts, each represented by a triangulation mesh with varying number of non-corresponding points and each with manufacturing and measurement noise. After applying the functional map method, each point $x$ on the $i$th mesh ${\mathcal M}_i$ is associated with a deviation measure $D^i(x)$, $i=1, 2, ..., m_0$. Note $D^i(x)$ and $D^j(y)$ tend to be correlated when $i=j$ and are not directly comparable when $i\neq j$, since $x\in{\mathcal M}_i$ and $y\in{\mathcal M}_j$ may be matched with different points on the CAD model. Such properties make it hard to utilize the in-control deviation measures $D^i(x)$ directly. On the other hand, the single threshold method circumvents this problem by considering the maximum deviation over each mesh, $D^i_{\max}\triangleq\max_{x\in{\mathcal M}_i} D^i(x)$, $i=1, 2, ..., m_0$. When $m_0$ is large enough, $D^i_{\max}$ forms an empirical distribution for acceptable parts and the $(\lfloor\alpha m_0\rfloor+1)$th largest $D^i_{\max}$ will act as the threshold, or the critical value. Points on the defective part whose deviation $D(x)$ exceeds this threshold are treated as significantly different from the corresponding points on the CAD model and are highlighted as local defects. The single threshold method is summarized in Algorithm \ref{algo3}.

\begin{algorithm}[h]
\SetAlgoLined
\KwIn{Defective part $\mathcal{A}$,  CAD design $\mathcal{B}$, and $m_0$ Phase I (in-control) parts ${\mathcal M}_i$, $i=1, 2, ..., m_0$}
\KwOut{A set $S$ of significantly defective points on $\mathcal A$}
\For{each Phase I part ${\mathcal M}_i$}{
    Apply Algorithm \ref{algo2} on ${\mathcal M}_i$ and $\mathcal B$, obtaining the point-to-point mapping $T_i: {\mathcal M}_i\to \mathcal B$\\
    For each point $x\in{\mathcal M}_i$, calculate its deviation from the CAD model $D^i(x)\triangleq\|\bf f_x-\bf g_{T_i(x)} \|$\\
    Record the maximum value $D^i_{\max}\triangleq\max_{x\in{\mathcal M}_i} D^i(x)$
    }
Take the $100(1-\alpha)$th percentile of the set $\{D^i_{\max}\}_{i=1}^{m_0}$ as the threshold, denoted as $D_{\text{thres}}$\\
Apply Algorithm \ref{algo2} on $\mathcal A$ and $\mathcal B$, obtaining the point-to-point mapping $T_A: \mathcal A\to \mathcal B$\\
For each point $x\in\mathcal A$, calculate its deviation from the CAD model $D^A(x)\triangleq\|\bf f_x-\bf g_{T_A(x)} \|$\\
Return $S=\{x\in\mathcal A|D^A(x)>D_{\text{thres}}\}$
\caption{The single threshold method}
\label{algo3}
\end{algorithm}

\cite{holmes1996nonparametric} prove that this single threshold method with the maximum statistic has both weak and strong control over the family-wise error rate (FWER), defined as the probability of making at least one false discovery in multiple hypothesis tests. A test procedure is said to have weak type I error control if FWER$\leq\alpha$ is guaranteed when all null hypotheses are true, and has strong type I error control if FWER$\leq\alpha$ holds for any possible combination of true and false null hypotheses. In our diagnostic problem, weak control means the highlighted area is false with probability at most $\alpha$ when no point-wise deviation is in fact significant, while a strong control guarantees a type I error of $\alpha$ for the highlighted area regardless of whether the other points are significant or not. These are desirable properties.

\begin{figure}
	\centering
		\includegraphics[scale=0.15]{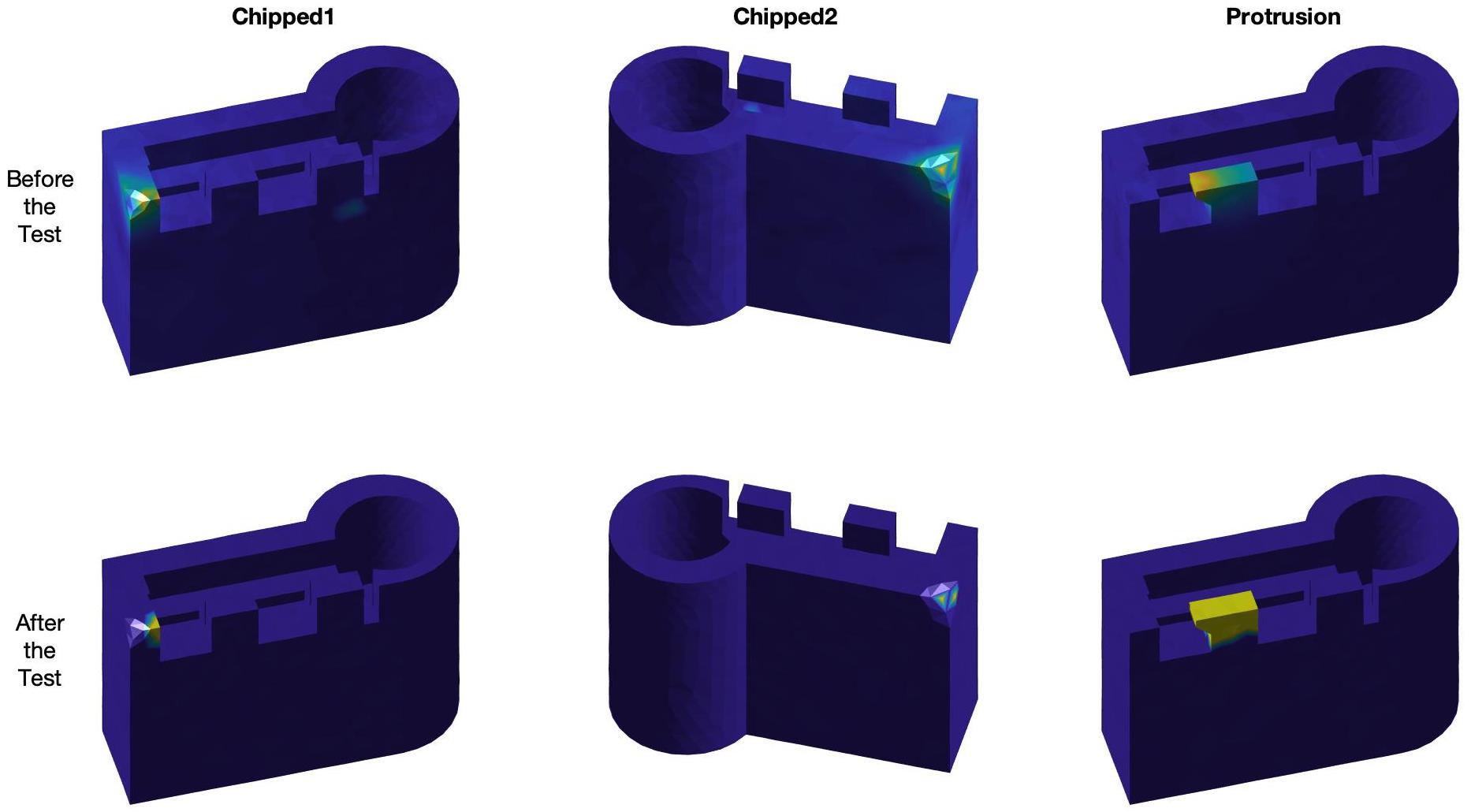}
	\caption{Defective parts with significant points highlighted in yellow. Top row: without applying the thresholding method. Bottom row: after applying the thresholding method. $m_0$=100, $\alpha=0.05$.}
	\label{SingleThres}
\end{figure}

Figure \ref{SingleThres} displays three random realizations for each of the defective parts previously shown in Figure \ref{highlight}. Using the notation in Algorithm \ref{algo3}, parts on the first row are color-coded by $D^A(x)$ and parts on the second row by $\delta_S(x)$, which equals 1 if $x\in S$ (set of points classified as significant defects) and 0 otherwise. Parts on the second row have highlighted the significant points in yellow and non-significant points in dark purple. The local defects in all three parts are correctly identified both with and without the single threshold method. However, in addition to accurately highlighting real defects, both the ``chipped1'' and ``chipped2'' parts have also small light blue areas (first row of parts), one slightly under the rightmost tooth and one on the top surface near the cylindrical region. They represent false detections caused by noise. Comparing with the corresponding parts on the second row, we can see how the thresholding algorithm is able to filter out such false alarms due to noise and to focus only on the true local defects. A second example showing the application of the threshold method is included in the Appendix.

\section{An algorithm for finding a region of interest where defects are likely to occur}\label{FM.ROI}

Sometimes a region of interest (ROI) can be defined by persons familiar with the part design, to indicate where  defects are expected to occur. Such information can be easily incorporated into the functional map framework. We can simply apply a filter on the point-wise dissimilarities $D^A(x)$, by defining
$$
D^A_\text{ROI}(x)=\begin{cases}
D^A(x), & \text{ if } x\in \text{ROI} \\
0, & \text{ otherwise }
\end{cases}
$$
For instance, the rightmost part on Figure \ref{highlightROI_zoom} shows a partial hex nut with a crack, colored by $D^A_\text{ROI}(x)$. As can be seen, the region whose high deviation is caused purely by the perturbation introduced by the boundary, marked by the red box on the third part, is successfully filtered out. The crack is correctly highlighted by focusing only on the region of interest. 

The ROI can not only help us better locate a defective region, it also reduces notably the computational complexity listed in Table \ref{compute}. Recall how in Algorithm \ref{algo2}, to perform step 10 for all points $x\in \mathcal A$ we need to calculate matrix multiplication $\tilde{\Phi}^BC'{\tilde{\Phi}^{A \prime}}$, where $\tilde{\Phi}^A$ and $\tilde{\Phi}^B$ are matrices consisting of the orthornormalized eigenvectors $\{\tilde{\phi}_i^A\}$ and $\{\tilde{\phi}_i^B\}$ as columns, respectively. Calculating $\tilde{\Phi}^BC'$ is an $\mathcal{O}(n_Bp)$ operation assuming $C$ diagonal, and multiplying the pre-calculated $\tilde{\Phi}^BC'$ by $\tilde{\Phi}^{A\prime}$ is the most computational expensive operation, $\mathcal{O}(n_An_Bp)$. This is because without additional information we need to evaluate $T_F(\delta_x)$ for all $x\in\mathcal A$. However, if we perform step 10 only for points in the ROI instead of the whole mesh $\mathcal A$, the computational complexity is reduced to $\mathcal{O}(n_Rn_Bp)$, where $n_R$ is the number of points in the region of interest and thus much smaller than the original $n_A$. 

A ROI, however can not always be defined a priori. For this reason, the following method automatically finds a ROI likely to contain a local defect when this cannot be defined a priori in practice. The main idea is to recursively partition the object into two connected components and select the component that deviates more from the CAD model, in a similar way to a binary search algorithm. Thus, the key to this method resides in finding a systematic method to partition the object and defining a measure to evaluate the deviation between each pairs of components. In what follows we explain these two steps.

Suppose we have identified, via SPC, a defective part $\mathcal A$ and have available its CAD design $\mathcal B$. To be able to accurately evaluate the shape deviation for a region of $\mathcal A$, we need to compare it against the corresponding region on $\mathcal B$. This means that when we partition $\mathcal A$, $\mathcal B$ needs to be partitioned in the same way. One approach to ensure this is to utilize the general shape of the two objects, which in manufacturing should be very close since we assume the defect only occurs in a small region on the defective part. We propose to use the {\em nodal domains} of the LB eigenvector corresponding to the first non-zero eigenvalue, which is usually the second eigenvector, given that meshes are typically connected. 

Let ${\mathcal A}_+$ (or ${\mathcal B}_+$) and ${\mathcal A}_-$ (or ${\mathcal B}_-$) be the nodal domains with positive and negative values for the second LB eigenvector of mesh $\mathcal A$ (or $\mathcal B$, respectively). We point out that despite the notation, ${\mathcal A}_+$ does not necessarily correspond to the same region as ${\mathcal B}_+$, since the sign of the eigenvectors is always ambiguous. To correctly match the components of $\mathcal A$ (namely ${\mathcal A}_+$ and ${\mathcal A}_-$) with the components of $\mathcal B$ (namely ${\mathcal B}_+$ and ${\mathcal B}_-$), we compare the number of points each component contains and switch the notations between ${\mathcal B}_+$ and ${\mathcal B}_-$ if necessary, an idea due to \cite{hamidian2019surface}. 

Now suppose we obtain four sub-meshes, with ${\mathcal A}_+$ corresponding to ${\mathcal B}_+$ and ${\mathcal A}_-$ corresponding to ${\mathcal B}_-$, respectively. The next step is to determine whether ${\mathcal A}_+$ deviates  from ${\mathcal B}_+$ more than what  ${\mathcal A}_-$ deviates from ${\mathcal B}_-$. The answer determines which pair of components we focus on in the next iteration. This problem goes back to the comparison of the shape between two meshes. Based on our previous results \citep{ZhaoEDC,ZhaoEDC-PE}, we use the first 15 eigenvalues of the LB operator estimated by the cubic FEM method as a shape feature for each component. That is, we compare $\sum_{i=1}^{15} |\lambda^A_{+,i}-\lambda^B_{+,i}|$ and $\sum_{i=1}^{15} |\lambda^A_{-,i}-\lambda^B_{-,i}|$, where $\lambda^A_{+,i}$ denotes the $i$th LB eigenvalue of sub-mesh ${\mathcal A}_+$ and we use similar notation for the other partitions. Finally, the pair of sub-meshes that results in a larger measure of this difference is selected for partition in the next iteration. This recursive method is summarized in Algorithm \ref{algo4}, where $|\mathcal{A}_+|$ is the cardinality or mesh size of $\mathcal{A}_+$.

\begin{algorithm}[h]
\SetAlgoLined
\KwIn{Defective part $\mathcal{A}$ and CAD design $\mathcal{B}$}
\KwOut{A region of interest $\mathcal A_R$ on $\mathcal A$}
\For{i=1:iter}{
    Partition $\mathcal{A}$ into $\mathcal{A}_+$ and $\mathcal{A}_-$ based on the sign of the second LB eigenvector, estimated using cubic FEM\\
    Partition $\mathcal{B}$ into $\mathcal{B}_+$ and $\mathcal{B}_-$ based on the sign of the second LB eigenvector, estimated using cubic FEM\\
    \uIf{ $(|\mathcal{A}_+|-|\mathcal{A}_-|)\cdot(|\mathcal{B}_+|-|\mathcal{B}_-|)<0$}
    {Set $\mathcal{B}_+$, $\mathcal{B}_-$ = $\mathcal{B}_-$, $\mathcal{B}_+$}
    Calculate the first 15 LB eigenvalues for $\mathcal{A}_+$, $\mathcal{A}_-$, $\mathcal{B}_+$, $\mathcal{B}_-$, respectively\\
    \uIf{$\sum_{i=1}^{15} |\lambda^A_{+,i}-\lambda^B_{+,i}|$ > $\sum_{i=1}^{15} |\lambda^A_{-,i}-\lambda^B_{-,i}|$} 
    {Set $\mathcal{A}=\mathcal{A}_+$ and $\mathcal{B}=\mathcal{B}_+$}
    \Else
    {Set $\mathcal{A}=\mathcal{A}_-$ and $\mathcal{B}=\mathcal{B}_-$}
    }
Return $\mathcal A_R=\mathcal A$
\caption{A recursive method to define a region of interest}
\label{algo4}
\end{algorithm}

\begin{figure}[h]
	\centering
		\includegraphics[scale=0.40]{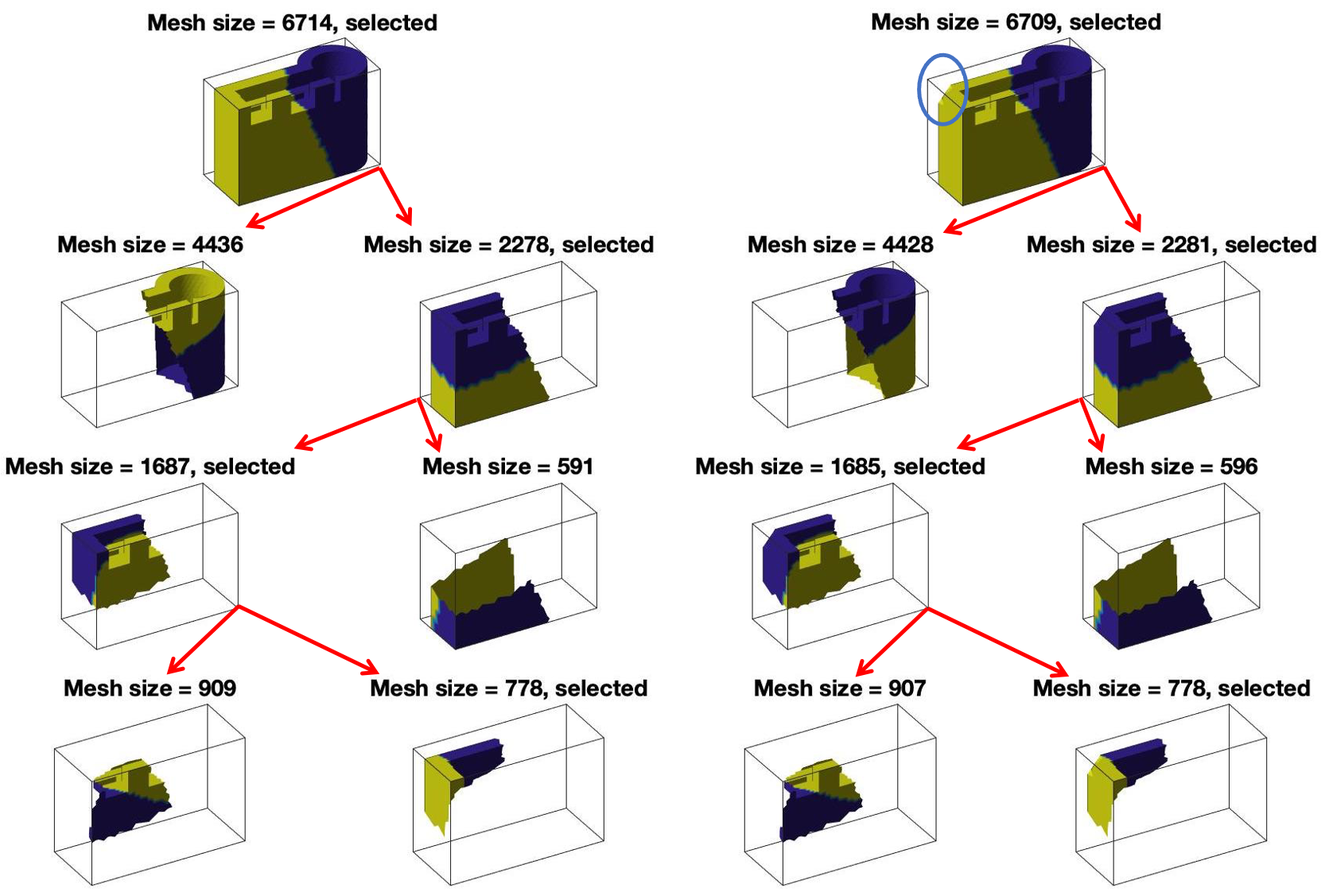}
	\caption{Recursive partitioning applied to the ``chipped2'' part to define a region of interest. Each row corresponds to an iteration, where starting from row 2, column 1 to 4 represent sub-meshes $\mathcal{B}_+$, $\mathcal{B}_-$, $\mathcal{A}_+$, and $\mathcal{A}_-$, respectively. The red arrows indicate how the selected pair of components are further partitioned in the next iteration. Colors indicate the sign of the second LB eigenvector (yellow for positive and purple for negative). The algorithm correctly matches the sub-meshes and selects the pair that contains the true local defect, circled in blue on the top row.}
	\label{binaryROI}
\end{figure}

Figure \ref{binaryROI} demonstrates how Algorithm \ref{algo4} is applied to define a ROI on the ``chipped2'' part (from Figure \ref{SingleThres}) and its CAD model. As the figure shows, the algorithm consistently partitions similar shapes (indicated by the colors), matching the four sub-meshes into two corresponding pairs (components in column 1 matched with those in 3 and components in column 2 matched with those in column 4), and selects the pair that contains the actual chipped corner. We point out that this recursion, however, cannot be applied indefinitely until the sub-meshes are small enough to only contain the local defect. In other words, this method cannot be used independently as a diagnostic scheme itself. The is because even though the partitions are roughly consistent for similar shapes, they cannot be {\em identical} due to small shape differences in the meshes and the different original mesh sizes. Thus, each application of the partitioning method introduces further differences into the newly obtained sub-meshes, which keep accumulating at each iteration. Eventually, the accumulated errors are large enough to disturb steps 4 and 7 in Algorithm \ref{algo4}, and the method will fail. Due to this reason, we propose using Algorithm \ref{algo4} for at most 2-3 iterations only to define  a ROI,  and then use the method in section \ref{FM.ROI} to localize in detail the defective region {\em within} the ROI. 

\section{Conclusions and further research}

We have presented a new defect localization method based on intrinsic differential properties of the manifolds under study that does not require registration. 
The method finds a point to point map between manifolds by first constructing a functional map between functions on each manifold. By using the Laplace Beltrami eigenvectors and a normalized version of the heat kernel signatures, the proposed method accurately matches points across two objects while being completely intrinsic. Therefore, our functional map method has consistently good performance regardless of the initial location and orientation of the scanned defective part, an important practical matter due to increasing popularity of hand-held non-contact sensors used in industrial settings. The stability of the functional map method makes it possible to infer the natural variability of the shape dissimilarities, allowing for statistical tests to find statistically significant shape deviations. A single threshold method was  introduced to handle the multiple comparison problem that arises from the massive number of simultaneous tests of hypothesis. 

Another advantage of our method over previous methods that require ICP-registration pre-processing is the computational benefits. Although both methods have an overall computational complexity of $\mathcal{O}(n_An_B)$ (with $n_X$ denoting the number of points in mesh $X$), our numerical experiments show that the functional map method scales much better with increasing mesh sizes. When a region of interest is defined, the functional mapping method naturally has a reduced complexity and for this reason we presented a new intrinsic recursive partitioning method to define a ROI where defects are likely to concentrate.

There are presently two limitations of our functional map method which we leave for further research. First, in common to all spectral Lapacian approaches, it may not perform well for parts with many symmetries. Symmetries create repeated eigenvalues and eigenspaces of dimensions higher than one, and this runs counter to the assumption of a diagonal $C$ transformation matrix in our method. While advanced manufactured parts with many symmetries are not common, it is still a matter of further research to find a modification of the method that overcomes the eigenvalue multiplicity problem, using, perhaps, intrinsic information other than the LB spectrum. Secondly, since the heat kernel signatures we utilize are affected by boundaries of an open surface, matrix $C$ may not be accurately estimated when matching two open meshes. 
How to overcome this problem, possibly by defining functions other than the HSK on each manifold, is also left for further research.

\noindent{\bf Supplementary materials}.- Matlab code that implements all algorithms and data files containing the meshes used in the examples are provided.

\bibliographystyle{informs2014} 
\bibliography{bibliography.bib}

\section*{Appendix. Technical and computational details.}

\begin{theorem} \label{LemmaC} The coefficients $\{c_{ij}\}$ fully determine the functional map $T_F$. 
\end{theorem}

{\em Proof}. We show this by considering an arbitrary real-valued function $f$ defined on $\mathcal{A}$. This function can be uniquely represented as a linear combination of the basis functions $f=\sum_i \alpha_i\phi_i^A$. The image of $f$ in $\mathcal{F}(\mathcal{B},\mathbb{R})$ is
\beeq
\label{Tf}
g\triangleq T_F(f)=T_F\left(\sum_i \alpha_i\phi_i^A\right)=\sum_i \alpha_i T_F(\phi_i^A)=\sum_i \alpha_i\sum_j c_{ij}\phi_j^B=\sum_j\left(\sum_i \alpha_ic_{ij}\right)\phi_j^B
\eeeq
where we used the property that $T_F$ is linear: 
$$T_F(\alpha_1 f_1+\alpha_2 f_2)=(\alpha_1 f_1+\alpha_2 f_2)\circ T^{-1}=\alpha_1 f_1\circ T^{-1}+\alpha_2 f_2\circ T^{-1}=\alpha_1 T_F(f_1)+\alpha_2 T_F(f_2).$$
When $f$ is known, $\alpha_i=\langle f, \phi_i^A \rangle$ is determined for all $i$, and therefore $g$ can be obtained provided the $c_{ij}$'s are known. Hence, $T_F$ is completely defined by the coefficients $\{c_{ij}\}$ in matrix $C\triangleq (c_{ij})$, and we can recover the point-to-point mapping $T$ by finding matrix $C$. $\blacksquare$

\subsection*{Intrinsic nature and independence from the mesh size of the normalized, scaled Heat Kernel Signature}

\begin{figure}
	\centering
		\includegraphics[scale=0.20]{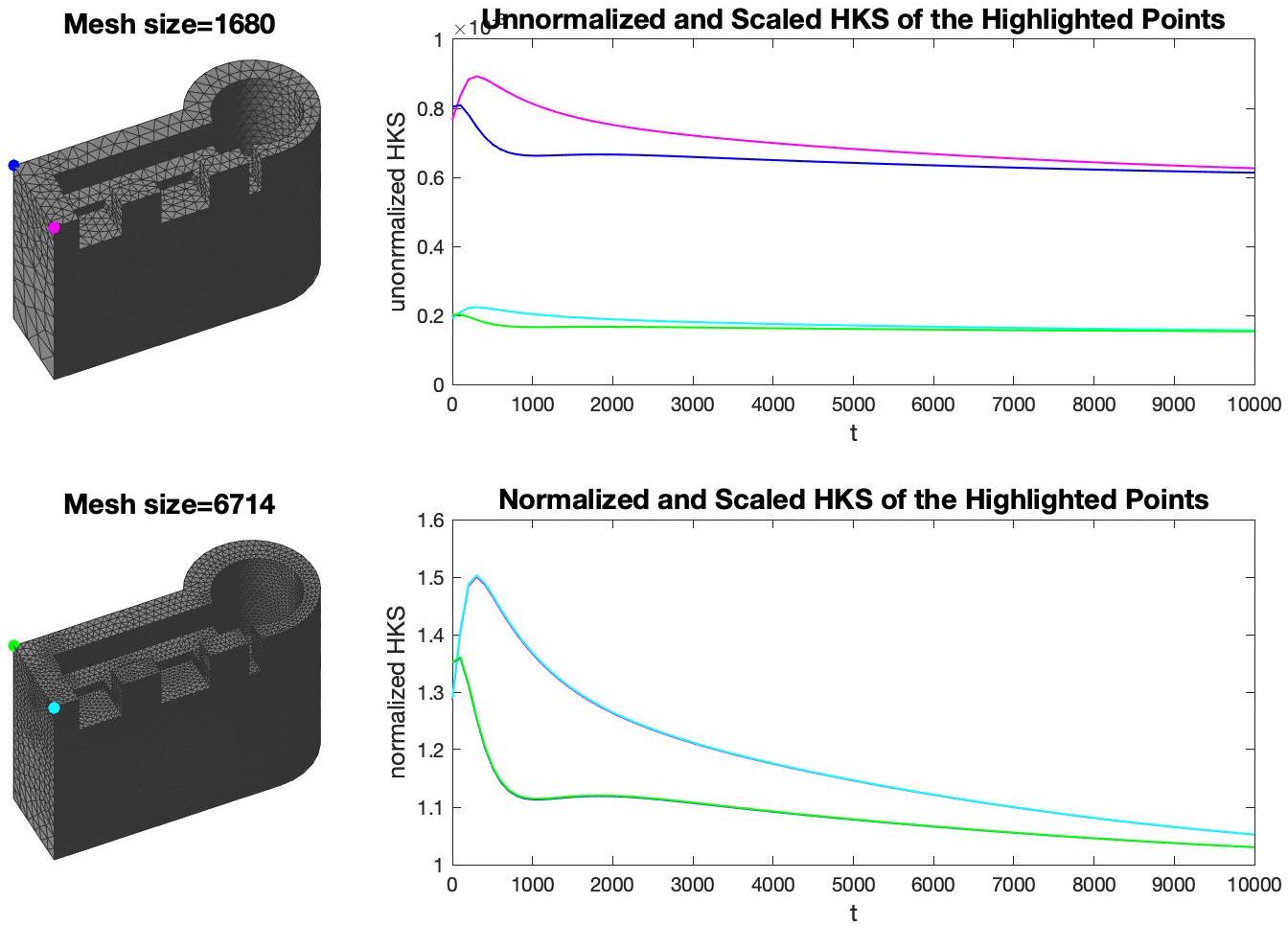}
	\caption{Scaled HKS and normalized, scaled HKS functions of four selected points on meshes of different sizes. The normalized, scaled HKS is independent of the mesh size.}
	\label{HKSplot}
\end{figure}

To demonstrate that the normalized, scaled HKS is intrinsic and independent of the mesh size being utilized, Figure \ref{HKSplot} plots the unnormalized and normalized scaled HKS versus time $t$ for four selected points, where a point and its corresponding HKS curve is drawn with the same color. The two prototype parts on the left have the same shape but considerably different mesh sizes. On the upper right, all four (unnormalized) scaled HKS curves have different shapes. However, on the bottom right, where the scaled HKS is normalized by the mesh size, points at the same location have nearly identical HKS regardless of the different mesh sizes, yet the normalized and scaled HKS is still able to distinguish points with different local shape information, hence there are only two functions visible corresponding to the different locations. We point out that the two parts in the figure are plotted with the same orientation for display purposes, but the scaled HKS remains the same even when the parts are differently oriented, since both the LB eigenvalues and LB eigenvectors that we use to calculate the scaled HKS are intrinsic, i.e., independent of the ambient space. In conclusion, the normalized and scaled HKS contains local shape information and is independent of mesh quality and rotation, providing correspondences across different scanned objects. 

\subsection*{Estimation of the $C$ matrix.}
\label{AppE1}
The functional map framework is based on solving the equation:
\beeq
B=C'A
\eeeq
where 
\beeq
B = \begin{pmatrix} \beta_{11} & \beta_{12} & \cdots & \beta_{1K} \\ 
\beta_{21} & \beta_{22} & \cdots & \beta_{2K} \\ 
\vdots & \vdots & \ddots & \vdots \\
\beta_{p1} & \beta_{p2} & \cdots & \beta_{pK}
\end{pmatrix},
\quad
A = \begin{pmatrix} \alpha_{11} & \alpha_{12} & \cdots & \alpha_{1K} \\ 
\alpha_{21} & \alpha_{22} & \cdots & \alpha_{2K} \\ 
\vdots & \vdots & \ddots & \vdots \\
\alpha_{p1} & \alpha_{p2} & \cdots & \alpha_{pK}
\end{pmatrix}
\eeeq
with $\beta_{jk}=\langle g_k, \phi_j^B\rangle$ and $\alpha_{ik}=\langle f_k, \phi_i^A\rangle$. By using the orthonormalized LB eigenvectors, matrix $C$ can be assumed to be diagonal, whose elements can be estimated using the least squares method. We minimize the squared Frobenius norm of $B-C'A$:
\beeq
\label{LS}
L_1 = ||B-C'A||^2 = \sum_{j=1}^p\sum_{k=1}^K (c_{jj}\alpha_{jk}-\beta_{jk})^2
\eeeq
Setting the partial derivative of $L_1$ to zero, we have
\beeq
\label{unconstrained}
\frac{\partial L_1}{\partial c_{jj}}=2\sum_{k=1}^K(c_{jj}\alpha_{jk}-\beta_{jk})\alpha_{jk}=0, \quad
c_{jj}^0=\frac{\sum_{k=1}^K\alpha_{jk}\beta_{jk}}{\sum_{k=1}^K\alpha_{jk}^2}=\frac{{\boldsymbol \alpha}^j{\boldsymbol \beta}^{j\prime}}{{\boldsymbol \alpha}^j{\boldsymbol \alpha}^{j\prime}}
\eeeq
where ${\boldsymbol \alpha}^j$ and ${\boldsymbol \beta}^j$ are the $j$th rows in A and B, respectively. In summary, the diagonal matrix $C$ can be estimated by $C=\text{diag}(AB')(\text{diag}(AA'))^{-1}$, where $\text{diag}(M)$ denotes a diagonal matrix consisting of only the diagonal elements of $M$.

According to Theorem 5.1 in \cite{ovsjanikov2012functional}, matrix $C$ should be orthonormal when the original point mapping $T$ is volume preserving, indicating that the elements should be either $1$ or $-1$ when $C$ is diagonal. In practice, the volume preserving property of $T$ will be violated due to mesh discretization, manufacturing and measurement noise, and the small shape difference between the $\mathcal A$ and $\mathcal B$ objects. Among all these factors, the effect of mesh discretization can be accurately quantified. We approach this by looking at the magnitude of the coefficients $\beta_{jk}$ and $\alpha_{ik}$. Take $\beta_{jk}=\langle g_k, \phi_j^B\rangle$ as an example, when the manifold $\mathcal B$ is discretized, $\beta_{jk}$ is numerically calculated as 
$$\beta_{jk}=\sum_{x\in\mathcal B} g_k(x)\phi_j^B(x)$$
Here $g_k(x)$ is independent of the mesh size after the scaled HKS is normalized, $\phi_j^B(x)$ is proportional to $1/\sqrt{n_B}$ since the eigenvectors are orthonormal, and there are $n_B$ terms to sum up. Hence, the magnitude of $\beta_{jk}$ is proportional to $\sqrt{n_B}$. Similarly, the magnitude of $\alpha_{ik}$ is proportional to $\sqrt{n_A}$. There are two options to discount this effect, either we can normalize $\beta_{jk}$ by $1/\sqrt{n_B}$ and $\alpha_{ik}$ by $1/\sqrt{n_A}$, respectively, or we adjust the theoretical value for the elements of $C$ from $\pm 1$ to $\pm \sqrt{n_B/n_A}$. These two options are essentially equivalent, and we choose option 1 to have a more uniform expression to estimate matrix $C$. 

\begin{figure}[h]
	\centering
		\includegraphics[width=\textwidth]{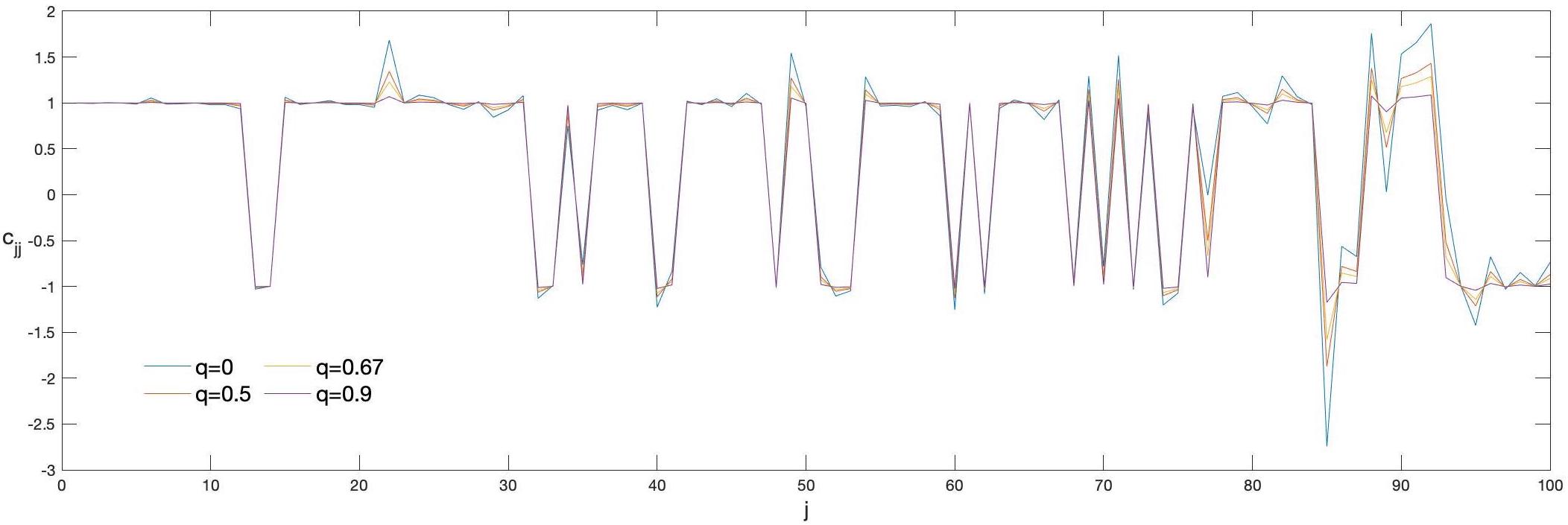}
	\caption{Values of the first 100 diagonal elements in the $C$ matrix with increasing values of $q$. As it can be seen, the larger the $q$, the closer the $C$ diagonal elements are to +1 or -1.}
	\label{ridgeplot}
\end{figure}

For the following derivations, we still expect the magnitude of the diagonal elements of $C$ to be approximately 1, but notice this will not be exactly true due to noise and small shape differences. This constraint can be incorporated in the above optimization process via Ridge Regression, penalizing deviations in the diagonals of $C$ away from magnitude 1. For this purpose, we introduce the Lagrange multipliers $\{\theta_j\}$ in the ridge-like objective function (\ref{LS}):
\beeq
\label{ridge}
L_2 = ||B-C'A||^2+\sum_{j=1}^p\theta_j(|c_{jj}|-1)^2 = \sum_{j=1}^p\sum_{k=1}^K (c_{jj}\alpha_{jk}-\beta_{jk})^2+\sum_{j=1}^p\theta_j(|c_{jj}|-1)^2
\eeeq
Taking the partial derivative with respect to $c_{jj}$:
\beeq
\begin{aligned}
\frac{\partial L_2}{\partial c_{jj}}&=2\sum_{k=1}^K(c_{jj}\alpha_{jk}-\beta_{jk})\alpha_{jk}+2\theta_j(|c_{jj}|-1)\text{sign}(c_{jj})\\
&=2\sum_{k=1}^K(c_{jj}\alpha_{jk}-\beta_{jk})\alpha_{jk}+2\theta_jc_{jj}-2\theta_j\text{sign}(c_{jj})
\end{aligned}
\eeeq
and setting $\frac{\partial L_2}{\partial c_{jj}}|_{c_{jj}^*}=0$, we obtain:
\beeq
\label{ridgeSol}
\begin{aligned}
c_{jj}^*&=\frac{\sum_{k=1}^K\alpha_{jk}\beta_{jk}+\theta_j\text{sign}(c_{jj}^*)}{\sum_{k=1}^K\alpha_{jk}^2+\theta_j}=\frac{{\boldsymbol \alpha}^j{\boldsymbol \beta}^{j\prime}+\theta_j\text{sign}(c_{jj}^*)}{{\boldsymbol \alpha}^j{\boldsymbol \alpha}^{j\prime}+\theta_j}\\
&=\frac{c_{jj}^0{\boldsymbol \alpha}^j{\boldsymbol \alpha}^{j\prime}+\theta_j\text{sign}(c_{jj}^0)}{{\boldsymbol \alpha}^j{\boldsymbol \alpha}^{j\prime}+\theta_j}=\begin{cases}
c_{jj}^0-(c_{jj}^0-1)\frac{\theta_j}{{\boldsymbol \alpha}^j{\boldsymbol \alpha}^{j\prime}+\theta_j}, & c_{jj}^0>0\\
c_{jj}^0-(c_{jj}^0+1)\frac{\theta_j}{{\boldsymbol \alpha}^j{\boldsymbol \alpha}^{j\prime}+\theta_j}, & c_{jj}^0<0
\end{cases}
\end{aligned}
\eeeq
Here it is reasonable to assume the regularization term will not change the sign of the elements in the $C$ matrix, so $\text{sign}(c_{jj}^*)=\text{sign}(c_{jj}^0)$. Let $q_j=\frac{\theta_j}{{\boldsymbol \alpha}^j{\boldsymbol \alpha}^{j\prime}+\theta_j}$, then $c_{jj}^*$ can be simplified to be:
\beeq
c_{jj}^*=
\begin{cases}
(1-q_j)c_{jj}^0+q_j, & c_{jj}^0>0\\
(1-q_j)c_{jj}^0-q_j, & c_{jj}^0<0\\
\end{cases}
\eeeq
It is clear that the ridge solution is simply a convex combination of the unconstrained solution (\ref{unconstrained}) and the theoretical solution (1 for positive elements and -1 for negative elements). For simplicity, we set $q_1=q_2=\cdots=q_p=q$ so that all elements in the $C$ matrix receive the same weight in this convex combination. The values of $\theta_j$ can be chosen accordingly to achieve the desired weight $q$. It is obvious $q\in[0,1)$. When $q=0$,  $\theta_j=0$, and there is no regularization, the resulting $c_{jj}$ is the same as $c_{jj}^0$ in equation (\ref{unconstrained}). As $q$ increases, more weight is given to the theoretical diagonal values of $\pm 1$, so the $c_{jj}$'s tend to have absolute values of 1, the result of large penalization. Figure \ref{ridgeplot} plots the first 100 elements in an example $C$ matrix with increasing values of $q$. It is evident that larger $q$ values ``push'' positive $c_{jj}$'s closer to 1 and negative $c_{jj}$'s closer to -1. Note the regularization effect will not change the sign of the $C$ elements, justifying our assumption that $\text{sign}(c_{jj}^*)=\text{sign}(c_{jj}^0)$ in eq (\ref{ridgeSol}). In our tests, we used $q=4/5=0.8$ as a compromise to incorporate the variability in the observations without violating the theoretical diagonal values of 1 too much.

\subsection*{Determining the $t$ Parameter in the heat kernel signature} \label{AppE2}
As discussed in Section \ref{FM.HKS} in the paper, we recommend using the normalized and scaled heat kernel signatures (\ref{scaledHKS}) as the known corresponding functions. As shown in Figure \ref{HKSplot}, the scaled HKS changes as $t$ increases, and its shape can be seen as a ``profile" or curve feature associated with the corresponding point $x$. To capture such pattern as much as possible, we want a wide range of $t$ values, but evidently this range needs to be finite. 

In practice, the summation in the HKS is truncated up to $p$, the number of LB eigenvectors we use, and so are the heat kernel signatures:
\beeq
k_t(x,x) = \sum_{i=0}^{\infty} e^{-\lambda_i t} \phi_i(x)^2
\approx \sum_{i=0}^{p} e^{-\lambda_i t} \phi_i(x)^2.
\label{HKS}
\eeeq
Such approximation is valid if $e^{-\lambda_i t}\phi_i(x)^2 \to 0$ for $i>p+1$ since $\lambda_i$ is large for large $i$. Nonetheless, this does not hold when $t$ is too small, making $\sum_{i=p+1}^{\infty} e^{-\lambda_i t} \phi_i(x)^2$
not negligible any more and the calculated HKS inaccurate. As $t$ decreases, the omitted term that increases the fastest is $e^{-\lambda_{p+1} t} \phi_{p+1}(x)^2$, bounded by $e^{-\lambda_{p} t} \phi_{p+1}(x)^2$ from above. Therefore, to prevent such approximation errors from being too large, we need to ensure 
$$e^{-\lambda_{p} t} \phi_{p+1}(x)^2\leq \varepsilon, \quad \forall x$$
where $\varepsilon$ is a pre-specified precision threshold. This leads to
$$t \; \geq \; \frac{\log(\phi_{p+1}(x)^2)-\log\varepsilon}{\lambda_p}, \quad \forall x \quad \mbox{or}$$
$$t_{\min}=\max_x \frac{\log(\phi_{p+1}(x)^2)-\log\varepsilon}{\lambda_p}= \frac{\max_x\log(\phi_{p+1}(x)^2)-\log\varepsilon}{\lambda_p}$$
Since $\phi_{p+1}(\cdot)$ is an eigenvector with expected norm of one, $\phi_{p+1}(x)^2$ is at most one, so $t_{\min}=-\log\varepsilon/\lambda_p$ is the smallest value that provides a faithful HKS. In our numerical computations, we chose $\varepsilon=10^{-4}$, which results in $t_{\min}=4\log10/\lambda_p$.

When $t$ is infinitely large, all terms in eq (\ref{HKS}) are almost zero except when $i=0$, so all heat kernel signatures have the following limit:
\beeq
\lim_{t\to\infty} k_t(x,x) = e^{-\lambda_0 t} \phi_0(x)^2=\phi_0(x)^2=\frac{1}{n}
\eeeq
where $n$ is the number of points on the manifold under study. The last equal sign holds because $\phi_0(\cdot)$ is the constant eigenvector corresponding to an eigenvalue of zero. Since all points have the same limit, the scaled HKS has almost no discrimination power when $t$ is large. Thus, we can safely stop increasing $t$ when the HKS is close enough to this limit for all points:
$$\begin{aligned}
t_{\max} &= \min_{t} \left\{t \Bigm\vert  \sum_{i=0}^{p} e^{-\lambda_i t} \phi_i(x)^2\leq \frac{1}{n}+\varepsilon, \quad \forall x\right\}\\
&= \min_{t} \left\{t \Bigm\vert  \sum_{i=1}^{p} e^{-\lambda_i t} \phi_i(x)^2\leq \varepsilon, \quad \forall x\right\}
\end{aligned}$$
Again, as $t$ increases, $e^{-\lambda_1 t}\phi_1(x)^2\leq e^{-\lambda_1 t}$ is the term that decreases the slowest, so $t_{\max}$ can be approximated by:
$$t_{\max} \approx \min_{t}\left\{t \Bigm\vert  e^{-\lambda_1 t}\leq \varepsilon\right\}=-\frac{\log\varepsilon}{\lambda_1}$$
Again by choosing $\varepsilon=10^{-4}$, we have $t_{\min}=4\log10/\lambda_1$. 

Due to the analysis above, we recommend using $t$ values in the range from $t_{\min}=4\log10/\lambda_p$ to $t_{\max}=4\log10/\lambda_1$. For all results presented in this paper, we use 100 values of $t$ uniformly sampled within this range in the log scale so that heavier sampling occurs for smaller values of $t$. We point out the same range is used by \cite{SunHKS2009} as well, yet how they derived these bounds is unclear from their paper, so our analysis presented here provides a principled justification. 

\subsection*{Global optimality of the functional map method compared to ICP-based solutions} 
A clear advantage of our functional map method compared with registration methods based on the ICP algorithm or any of its many variants, is that, being intrinsic and invariant with respect to rigid transformations, the functional map is independent of the initial orientation and position of the part we want to match with the CAD model. The registration diagnostic based on the well-known ICP algorithm \citep{Besl}, on the other hand, relies heavily on the initial location/orientation of the part. This is not only a theoretical advantage: with the increasing use of high-quality hand-held scanners in industry, widely different orientations are possible. For example, Figure \ref{vsICP} in the paper shows two scenarios for matching the ``chipped1'' part with an ``acceptable'' part (see Figure \ref{plotTf}). The left most column plots the initial orientation of the ``chipped1'' part. Next to it is the acceptable' part in grey together with the ``chipped1'' part, transformed according to the ICP solution, in blue. The last three columns on the right color-code the ``chipped1'' part by the different point-wise deviations, calculated using ICP, functional mapping with cubic FEM, and functional mapping with linear FEM, respectively. Lighter colors indicate larger deviations and possible local defects. Each row corresponds to an initial orientation plotted in the first column. The second row shows that, starting with a ``bad'' initial orientation, ICP is trapped in a local optimum and fails to correctly align the two objects, which results in the inaccurate point-wise deviations as shown in the third column, where non-defective areas are falsely highlighted due to inflated error. On the other hand, our functional mapping method works uniformly well for all initial orientations, using either the cubic FEM or the linear FEM to approximate the Laplace-Beltrami eigenpairs, as evidenced in the last two columns.

\subsection*{An additional example on the thresholding approach to multiple point to point comparisons}

\begin{figure}
	\centering
		\includegraphics[width=\textwidth]{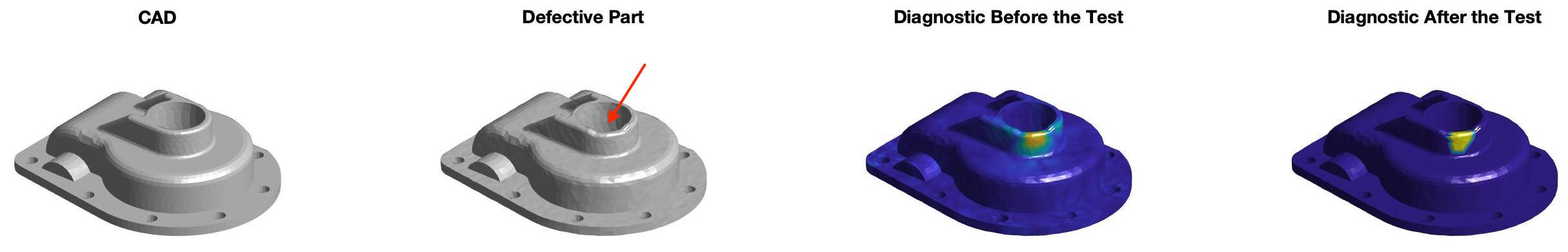}
	\caption{From left to right: the CAD design of a casted part, a defective part with a notch on the top of the cylindrical edge (see the red arrow), the defective part color-coded by $D^A(x)$, and the defective region identified by the single threshold method (in yellow).	}
	\label{SingleThresCast}
\end{figure}

Figure \ref{SingleThresCast} shows an example of a casted part which has a defect in the form of a notch on the top of the cylindrical edge. The two plots on the right are the diagnostic before and after the single threshold method is applied, respectively, where the lighter (yellow and green) regions are identified to be local defects. As we can see from the figure, although the functional map method can successfully locate the error by itself, the application of the single threshold method is able to further narrow down the particular defective region.

\end{document}